\documentclass[11pt,hidelinks]{article}

\usepackage{amsmath,amssymb,amsfonts, braket}
\usepackage{graphicx}
\usepackage{bm}
\usepackage{hyperref}
\usepackage[T1]{fontenc}
\usepackage{multirow}
\usepackage{hhline}
\usepackage{booktabs}
\usepackage{makecell}
\usepackage{romannum}
\usepackage[margin=0.75in]{geometry}

\usepackage{multibib}

\renewcommand{\figurename}{\textbf{Fig.}}
\renewcommand{\thefigure}{\textbf{\arabic{figure}}}%

\newcommand{\er}{Er$^{3+}$}
\newcommand{\ercawo}{Er$^{3+}$:CaWO$_4$}
\newcommand{\cawo}{CaWO$_4$}

\newcommand{\upg}{\uparrow_\text{g}}
\newcommand{\downg}{\downarrow_\text{g}}
\newcommand{\upe}{\uparrow_\text{e}}
\newcommand{\downe}{\downarrow_\text{e}}

\title{Spin-photon entanglement of a single \er{} ion in the telecom band}

\author{Mehmet T. Uysal$^1$\thanks{These authors contributed equally to this work.},  \L{}ukasz Dusanowski$^{1*}$, Haitong Xu$^{1*}$, Sebastian P. Horvath$^1$, Salim Ourari$^1$, \\ Robert J. Cava$^2$, Nathalie P. de Leon$^1$, Jeff D. Thompson$^{1}$\thanks{jdthompson@princeton.edu} }

\date{
	$^1$Department of Electrical and Computer Engineering, Princeton University,\\ Princeton, NJ, 08544, USA \\ 
 	$^2$Department of Chemistry, Princeton University, Princeton, NJ 08544, USA \\
}
\begin{document}

\maketitle



\maketitle

\textbf{Long-distance quantum communication using quantum repeaters is an enabling technology for secure communication, distributed quantum computing and quantum-enhanced sensing and metrology \cite{Awschalom2021,gisin2002quantum}. As a building block of quantum repeaters, spin-photon entanglement has been demonstrated with both atomic \cite{blinov2004observation,matsukevich2004quantum, chou2005measurement, volz2006observation,wilk2007single} and solid-state qubits \cite{gao2012observation,de2012quantum,Togan2010,Bernien2013,Bhaskar2020,lago2021telecom,ruskuc2024scalable}. However, previously demonstrated qubits with long spin coherence do not directly emit photons into the low-loss telecom band that is needed for long-distance communication. Here, we demonstrate spin-photon entanglement using a single \er{} ion in a solid-state crystal, integrated into a silicon nanophotonic circuit. Direct emission into the telecom band enables an entanglement rate of 1.48 Hz over 15.6 km of optical fiber, with a fidelity of 73(3)\%. This opens the door to large-scale quantum networks based on scalable nanophotonic devices and many spectrally multiplexed \er{} ions~\cite{Chen2020}.}

Individually addressable solid-state defects integrated with nanophotonic devices are particularly attractive for combining a single photon source and quantum memory in a scalable platform \cite{sipahigil2016integrated,jung2019spin,mouradian2017rectangular,kuruma2021coupling,rugar2021quantum,chakravarthi2023hybrid,ruskuc2024scalable}. However, a challenge is that the typical energy scale for atomic transitions (400 -- 900~nm) is outside the low-loss telecom band in standard optical fibers (1.5~$\mu$m), resulting in significant losses for long distance transmission.
Two approaches have been demonstrated to circumvent this challenge: single-photon frequency conversion \cite{de2012quantum, tchebotareva2019entanglement,krutyanskiy2019light,krutyanskiy2023telecom,van2020long,van2022entangling,knaut2024entanglement} and generating entangled photon pairs using spontaneous parametric downconversion (SPDC), where one photon is in the telecom band and the other matches the wavelength of an atomic memory~\cite{lago2021telecom,lago2023long}. However, these approaches suffer from limited efficiency and added noise, and in the case of SPDC, involve non-deterministic sources. Recently, direct telecom spin-photon entanglement of a single emitter has been demonstrated in an InAs/InP quantum dot \cite{laccotripes2023spin}, but direct entanglement with a long-lived spin suitable for long-distance transmission has not been reported.

The rare earth ion (REI) \er{} provides a direct spin-photon interface in the telecom band, but exploiting individual REIs as single photon sources and quantum memories is challenging because of their low photon emission rate and correspondingly increased sensitivity to decoherence. Recent work has demonstrated incorporating REIs into nanophotonic cavities to enhance the emission rate up to 1000-fold \cite{Dibos2018, Zhong2018, yu2023frequency, huang2023stark, ji2024nanocavity, ulanowski2022spectral, Yang2022, horvath2023strong}, enabling single-shot readout~\cite{Raha2020,Kindem2020,gritsch2024optical} and manipulating nearby nuclear spin ancillae~\cite{Kornher2020,Ruskuc2022,uysal2023coherent}, and remote entanglement of two Yb$^{3+}$:YVO$_4$ defects~\cite{ruskuc2024scalable}. Careful selection and engineering of host materials has also led to improved coherence properties. In particular, \ercawo{} has demonstrated coherence times of 23 ms in small ensembles \cite{LeDantec2021}, and low enough spectral diffusion to enable indistinguishable single photon generation in a Si nanophotonic cavity \cite{ourari2023indistinguishable}. However, despite these improvements, maintaining the joint spin and optical coherence during photon emission remains a challenge.

Here, we demonstrate spin-photon entanglement directly in the telecom band, with a single \er{} ion in \cawo{} integrated into a silicon nanophotonic device. The key advance enabling this demonstration is a spin-photon entanglement protocol capable of refocusing spin decoherence while waiting for a photon to be emitted; this protocol has the additional benefit of mitigating an unanticipated optically-induced spin dephasing mechanism. Through the large Purcell factor of the cavity ($P=342$), the absence of losses from frequency conversion, and the low propagation losses at the emission wavelength of $\lambda=1532.6$ nm, we reach a spin-photon entanglement rate of 1.48 Hz over 15.6~kilometers of optical fiber. This rate improves on previously demonstrated telecom band spin-photon entanglement using frequency converted solid-state sources \cite{de2012quantum, tchebotareva2019entanglement, knaut2024entanglement} and is comparable to frequency-converted spin-photon entanglement from trapped ions and single atoms in optical tweezers \cite{krutyanskiy2019light,krutyanskiy2023telecom,van2022entangling}. The fidelity is $F=0.73(3)$, beyond the classical bound of 0.5 \cite{Togan2010}. This opens the door to implementing long-distance quantum networks with \er{} devices.

\begin{figure*}[ht]
    \includegraphics[width= \textwidth]{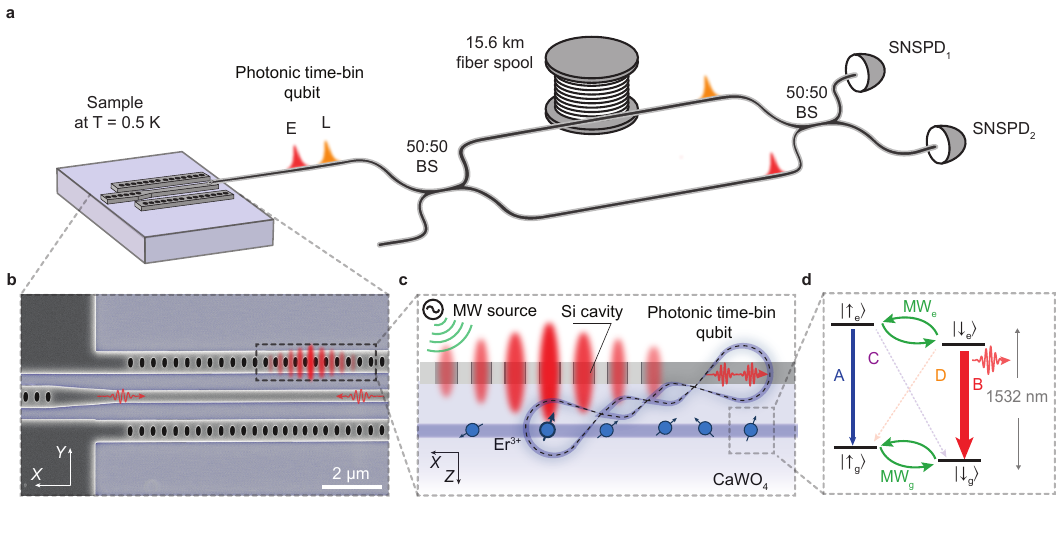}
    \caption{\textbf{Er$^{3+}$:CaWO$_4$ spin-photon interface.} \textbf{a} Optical setup used to generate and characterize spin-photon entanglement.
    Single photons entangled with the \er{} ion and encoded in early (E) or late (L) time bins pass through a Mach-Zehnder interferometer (MZI) with a 15.6 km fiber spool in one arm, and are detected at two superconducting nanowire single-photon detectors (SNSPDs).
    \textbf{b} Scanning electron microscope image showing a silicon nanophotonic cavity coupled to a bus waveguide, with a schematic of the resonant mode depicted in red. \textbf{c} Implanted erbium ions evanescently couple to the silicon photonic crystal cavity placed on the surface of the CaWO$_4$, and their spin transitions are driven by a microwave field. \textbf{d} Er$^{3+}$ level structure in an applied magnetic field with four optical (A-D) and two microwave (MW$_e$, MW$_g$) transitions. The cavity resonance frequency is chosen to match the $B$ transition, resulting in a larger Purcell enhancement for $\ket{\downarrow_e}$ ($\Gamma_B/\Gamma_A = 4.6$).
}
    \label{fig:Fig1}
\end{figure*}

Our experimental approach is depicted schematically in Fig.~\ref{fig:Fig1}a. The nanophotonic device architecture has previously been described in Refs.~\cite{ourari2023indistinguishable, Chen2021a}. Briefly, a hybrid structure is formed by a Si nanophotonic device bonded on the \cawo{} substrate (Fig.~\ref{fig:Fig1}b). The sample is cooled to $T=500$ mK in a $^3$He cryostat. Light is extracted from the cavity into a bus waveguide and ultimately an optical fiber though a grating coupler, and microwaves to drive spin transitions are applied using a nearby antenna~\cite{Chen2021a}. \er{} ions are implanted $\sim$10 nm below the surface of the \cawo{} crystal, within the evanescent field of the Si cavity, resulting in a strong enhancement of the emission rate (Fig.~\ref{fig:Fig1}c). Through careful control of the implantation and annealing conditions, we have observed single ions with spectral diffusion linewidths below 200~kHz and spectral diffusion less than the radiative linewidth on the timescale of photon emission~\cite{ourari2023indistinguishable}. The ion used in this work has a slightly broader spectral diffusion linewidth of 470~kHz (Extended Data Fig.~\ref{fig:SD}), and a radiative lifetime of $\tau_r = 18.4$~$\mu$s for the B transition that is resonant with the optical cavity (Fig.~\ref{fig:Fig1}d), given by the Purcell factor of $P=342$.

\begin{figure*}[ht]
	\centering
    \includegraphics[]{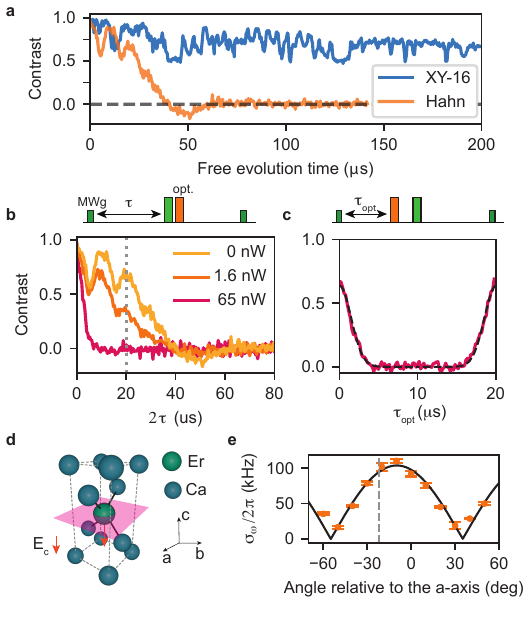} 
    \caption{\textbf{Optically induced spin dephasing.} \textbf{a} In the absence of optical illumination, the Hahn echo spin coherence is 30~$\mu$s, which can be extended to over 200~$\mu$s with an XY-16 sequence. \textbf{b} Applying an optical pulse in the middle of a Hahn echo sequence degrades the coherence. The effect is shown for 200 ns pulses at two optical powers: 1.6 nW, corresponding to the power used for optical $\pi$ pulses in the following experiments, 65 nW, to more clearly show the effect. \textbf{c} In a fixed sequence length ($2 \tau = 20\,\mu$s), the Hahn echo contrast is maximized by placing the optical pulse at the beginning or the end of the sequence, indicating that the optically induced dephasing can be refocused. The dashed line is a fit to a Gaussian decay envelope, $e^{-(\tau_{\textrm{opt}}/T_d)^2}$ with $T_d$ = 2.1~$\mu$s.
    \textbf{d} The four nearest neighbor calcium ions in the \ercawo{} unit cell (highlighted by black lines) show $S_4$ symmetry, invariant under a 90$^{\circ}$ improper rotation. An electric field along the $c$-axis (red arrows) lowers the symmetry, which lifts the in-plane g-tensor degeneracy and results in a linear shift of  the spin splitting~\cite{mims1965electric}. W and O atoms are not displayed.
    \textbf{e} The induced frequency fluctuation $\sigma_{\omega} = \sqrt{2}/T_d$ varies with the magnetic field orientation in the $aa$-plane as $\sigma_{\omega}(\phi) \propto |\sin(2\phi_B-2\phi_0)|$ with $\phi_0 = -35(1)^{\circ}$, confirming that it arises from a fluctuating electric field. The dashed line corresponds to the field orientation used in the spin-photon experiment, which is needed to obtain a cycling transition for this ion~\cite{Raha2020}.
    }
    \label{fig:Fig2}
\end{figure*}

In the standard time-bin spin-photon entanglement protocol~\cite{barrett2005}, optical excitation pulses are embedded in a Hahn echo sequence with a free precession period that is long enough to allow the excited state to fully decay between pulses in order to preserve spin-photon correlations. There are two challenges to implementing this directly in our \ercawo{} devices. First, maintaining coherence for significantly longer than $\tau_r$ requires faster dynamical decoupling than a Hahn echo: while the XY-16 coherence time is 200~$\mu$s, the Hahn echo coherence time is only 30~$\mu$s (Fig.~\ref{fig:Fig2}a), likely limited by fluctuating magnetic field noise from paramagnetic impurities \cite{ourari2023indistinguishable}. Second, the optical pulses themselves cause spin decoherence through a previously unreported mechanism, which is not refocused in the standard sequence.

We first investigate spin decoherence caused by optical pulses.
We discovered that the insertion of an optical pulse during free precession can reduce the Hahn echo coherence time significantly, to less than 4~$\mu$s (Fig.~\ref{fig:Fig2}b).
The effect does not depend on the optical pulses being resonant with the optical transition of \er.
The coherence is restored if the optical pulse is applied at the beginning or end of the echo sequence, which suggests a mechanism of spin spectral diffusion induced by the optical pulse (Fig.~\ref{fig:Fig2}c). Because of its $S_4$ site symmetry (Fig.~\ref{fig:Fig2}d), the spin precession frequency of \ercawo{} is linearly sensitive to both magnetic and electric fields, where the latter arises from a distortion in the electronic wavefunction that changes the magnetic moment~\cite{mims1965electric,LeDantec2021}. We disentangle their effects by measuring the induced spectral diffusion, $\sigma_{\omega}$, for different magnetic field orientations in the $ab-$plane, and observe a dependence of $\sigma_{\omega} \propto |\sin\left(2\phi-2\phi_0\right)|$ with $\phi_0 = 35^{\circ}(1)$ (Fig.~\ref{fig:Fig2}e), matching the previously measured electric field induced magnetic moment shift for \ercawo{}~\cite{mims1965electric}. 
Therefore, we attribute the optically-induced dephasing to charge noise, and based on the measured scaling with power and pulse duration, conclude that it results from a single-photon absorption process [Extended Data Fig.~\ref{fig:OptDep}]. 
We note that previous measurements of spin coherence in \ercawo{} were performed without optical illumination, and are therefore not sensitive to this effect \cite{ourari2023indistinguishable,wang2023single,LeDantec2021}.

\begin{figure*}[ht]
    \centering
    \includegraphics[]{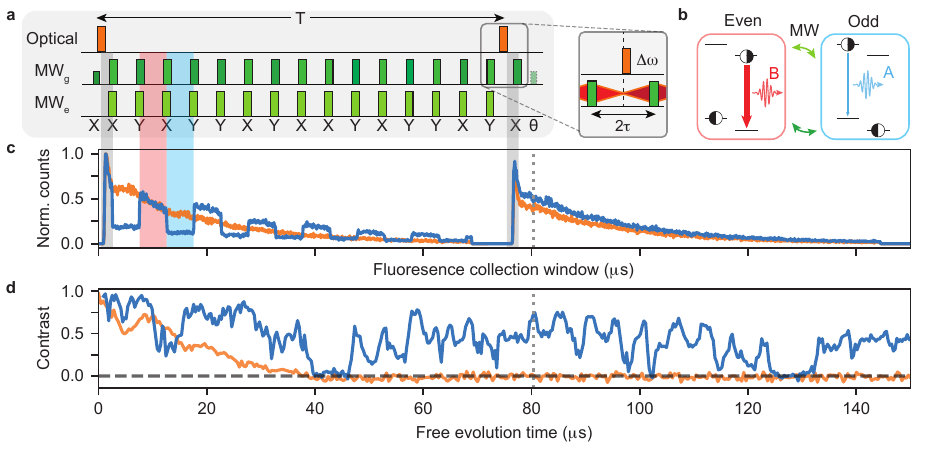} 
    \caption{\textbf{Entanglement protocol with dynamical decoupling.}
    \textbf{a} The pulse sequence for spin-photon entanglement. Two optical $\pi_B$ pulses interleaved with an XY-16 sequence on the MW$_g$ and MW$_e$ transitions, where the optical pulses are separated by 15 $\pi$ pulses on the spin. The decoupling sequence ends shortly after the heralding window (grey highlight) and the last excited state $\pi$ pulse also omitted since there is no need to continue decoupling beyond this point.
    A generalized form of the sequence is provided in Extended Data Fig.~\ref{fig:decoupling}.
    The last $\pi/2$ pulse on the MW$_g$ transition chooses the spin measurement basis. 
    The optical pulses are applied at refocusing points of the decoupling sequence, so that any random frequency shift of the spin, $\Delta\omega$, induced by the pulse is cancelled.
    \textbf{b} The emitter state before spontaneous emission transitions between the $\ket{\psi_E} = (\ket{\upg} + \ket{\downe})/\sqrt{2}$ (red highlight) and $\ket{\psi_O} = (\ket{\downg} + \ket{\upe})/\sqrt{2}$ (blue highlight) states for even and odd windows.
    \textbf{c} 
    Fluorescence during the standard protocol (orange) and the modified protocol (blue) is shown. For the modified protocol, the fluorescence switches between emission rates $\Gamma_B$ and $\Gamma_A$ with each $\pi$ pulse, corresponding to the states in panel b. The higher count rate in the very beginning of the sequence is due to a temporarily higher bias current when the SNSPD is biased on.
    \textbf{d} Ground manifold spin coherence relevant for entanglement attempts is measured for XY-16 (blue) and Hahn echo (orange), perturbed by an off-resonant optical pulse with equivalent power to excitation pulses (1.6~nW). The XY-16 total evolution time used in the entanglement generation is 80.5~$\mu$s, also indicated by the grey dashed line. 
    }
    \label{fig:Fig3}
\end{figure*}

To mitigate both the optically-induced spin dephasing and magnetic field noise, we introduce a modified spin-photon entanglement protocol with fast dynamical decoupling.
In the modified protocol, we apply an XY-16 decoupling sequence with a $\pi$ pulse spacing that is much shorter than the excited state lifetime. To preserve spin-photon correlations, we apply the $\pi$ pulses to both the ground and excited state spin manifolds simultaneously (Fig.~\ref{fig:Fig3}a,b). We apply two optical pulses separated by an odd number of $\pi$ pulses to create the spin-photon entangled state $(\ket{\downg}\ket{E}+\ket{\upg}\ket{L})/\sqrt{2}$, where $\ket{E}$ and $\ket{L}$ denote photons in the early and late time-bins, respectively. 
Since the optical pulses are applied at refocusing points of the XY-16 sequence, the optically induced spin dephasing is also refocused.

In Fig.~\ref{fig:Fig3}c, we compare the photon emission timing and spin coherence under optical illumination for both protocols. In a standard Hahn echo time-bin entanglement protocol, the spin coherence decays on the timescale of 20~$\mu$s, which is comparable to $\tau_r$. In the modified protocol, the spin coherence is protected beyond $150$~$\mu$s, provided the precise pulse spacing is chosen to avoid resonant interactions with the $^{183}$W nuclear spin bath.

\begin{figure*}[ht]
    \centering
    \includegraphics[]{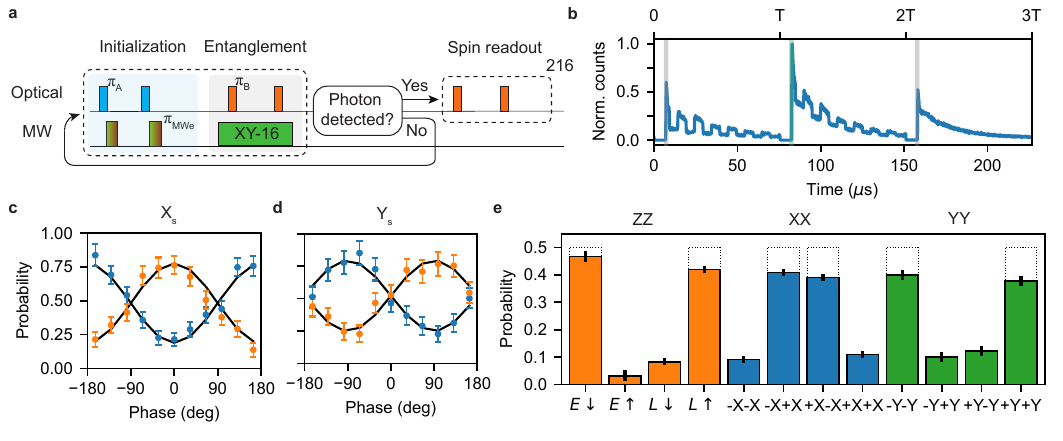}
    \caption{\textbf{Spin-photon entanglement generation.} \textbf{a} The experimental sequence used for spin-photon entanglement, showing that a spin-readout is only performed if a photon is detected. Otherwise the spin is re-initialized for another entanglement attempt.
    For efficient initialization, we use two optical $\pi_A$ pulses, followed by chirped excited state $\pi$ pulses in the excited state with high inversion, and perform readout with repeated optical $\pi_B$ pulses.
    \textbf{b} Fluorescence from the emitter after traveling through the MZI. The grey shaded regions mark the heralding window for the $Z_p$ photon basis measurement, while the green shaded region marks the heralding window for the $\Phi_p$ photon basis. 
    \textbf{c,d} Parity oscillations of the Bell state obtained by measuring the photonic qubit in the $\ket{+\phi}$ (orange) or $\ket{-\phi}$ (blue) state, corresponding to detection at SNSPD$_1$ or SNSPD$_2$, for spin-readout in the $X_s$ (c) or $Y_s$ basis (d).
    \textbf{e} Bell state tomography of the spin-photon entangled state measured in the ZZ basis (left), XX basis (center), and YY basis (right) giving a spin-photon entangled state fidelity of $F = 0.73(3)$. Dotted bars show ideal values and error bars correspond to one s.d.
    Note that the spin-photon correlations are corrected for the readout fidelity [Methods~S\ref{sec:init_and_readout}].}
    \label{fig:Fig4}
\end{figure*}

We generate and measure spin-photon entanglement using the sequence in  Fig.~\ref{fig:Fig4}a. To measure the photonic state, we send the emitted photons to an unbalanced Mach-Zehnder interferometer (MZI).
The time-bin encoded photon, with bins separated by $T=75.5$~$\mu$s, is split between a long arm of 15.6 km and a short arm (attenuated for equivalent loss) and re-combined on a 50:50 beam-splitter (BS) before going to two superconducting nanowire single photon detectors (SNSPD)(Fig.~\ref{fig:Fig1}a).
The detection of a single photon in the early ($[0,T]$) or late ($[2T,3T]$) time bins corresponds to a measurement of the photon in the $Z_p$ basis ($\ket{E}$,$\ket{L}$), while the detection of a photon in the central time bin, $[T,2T]$, corresponds to a measurement in the superposition basis $\Phi_p$, $\ket{\pm \phi}=\ket{E} \pm e^{-i\phi}\ket{L})/\sqrt{2}$, where $\phi$ is the relative phase accumulated in the interferometer (Fig.~\ref{fig:Fig4}b). Instead of stabilizing $\phi$, we track it continuously throughout the experiment using a reference laser sent through the same interferometer, resulting in a uniformly distributed set of measurement bases [Methods~S\ref{sec:phase_monitoring}].
After each entanglement attempt, the spin state is measured only if a photon is detected within the heralding window, using real-time control with a complex programmable logic device (CPLD) [Methods S\ref{sec:cond_op}]. 
We compute the entanglement fidelity from the visibility of the entanglement in the X, Y and Z bases (Fig.~\ref{fig:Fig4}c,d,e), $F=(1+E_X+E_Y+E_Z)/4$ \cite{tchebotareva2019entanglement}, where $\langle X_s \Phi_p \rangle = E_X\cos(\phi)$, $\langle Y_s \Phi_p \rangle = E_Y\sin(\phi)$ and $E_Z = \langle Z_s Z_p \rangle$. Here, $\Phi_p = \ket{+\phi}\bra{+\phi}-\ket{-\phi}\bra{-\phi}$. The experimentally observed visibilities are $\{E_X,E_Y,E_Z\} = \{0.60 (3),0.55 (3), 0.77 (5)\}$ [Methods~S\ref{sec:fidelity_estimation}], corresponding to a total fidelity of $F=0.73(3)$. The rate of successful entanglement generation is $R=1.48$~Hz. We note that we only accept photons emitted within the first 2.5~$\mu$s of the collection window, comprising 9.1\% of all emitted photons, to minimize sensitivity to errors from fast dephasing of the optical transition~\cite{metz2008effect} [Methods S\ref{sec:extended_spin_photon}].

The measured fidelity is in reasonable agreement with estimated visibilities of $E_{X/Y} = 0.63$ and $E_Z = 0.80$ based on independently measured sources of error [Methods~S\ref{sec:error_model}]. In particular, we identify significant contributions from spin decoherence and MW pulse errors ($\epsilon_{X/Y} = 25$\%, $\epsilon_{Z} = 14$\%), optical decoherence (7\%), background counts (5\%), initialization errors (3\%), and overlap of emission from the early and late time bins (2\%).

In this work, we have demonstrated spin-photon entanglement of a single \er{} ion directly in the telecom band over 15.6~km of optical fiber, enabled by our fast dynamical decoupling protocol that protects spin coherence during photon emission and mitigates optically induced dephasing. 
The demonstrated entanglement rate is comparable to that obtained with frequency-conversion from single trapped atoms or ions \cite{krutyanskiy2019light,krutyanskiy2023telecom,van2022entangling} and is higher than that obtained with frequency-converted solid-state sources \cite{de2012quantum,tchebotareva2019entanglement,knaut2024entanglement}.
The high rate for this distance is achieved by avoiding frequency conversion, and is not limited by the relatively long optical lifetime of the \er{} ion: after cavity enhancement, the lifetime is shorter than the photon propagation time through a long fiber link such as the one used here. 
We estimate that the entanglement rate can be increased to 150 Hz with improvements in optical and spin coherence to make use of all emitted photons ($\times11$), improvements to the cavity impedance matching and fiber coupling efficiency ($\times 6$)~\cite{tiecke2015efficient}, and faster spin reset ($\times1.5$) [Methods~S\ref{sec:entanglement_rate}]. The entanglement fidelity is mainly limited by coherence and dark counts, and should improve with the same enhancements.

These results pave the way to future long-distance quantum repeaters based on \er{} ions interfaced with scalable nanophotonic devices. While the long-term spectral diffusion is still significantly larger than the radiative linewidth, it was recently demonstrated that quasi-static depehasing can be refocused in the context of spin-spin entanglement protocols~\cite{ruskuc2024scalable,uysal2024rephasing}. A particularly exciting direction is to use multiplexed control of many emitters in the same cavity~\cite{Chen2020} to store multiple Bell pairs, which can enable higher-rate operation with reduced memory time requirements in a repeater protocol~\cite{collins2007multiplexed}, as well as entanglement distillation~\cite{bennett1996purification,kalb2017entanglement}.

\textbf{Acknowledgements} We acknowledge helpful conversations with Shimon Kolkowitz, Rose Ahlefeldt and Adam Turflinger. This work was primarily supported by the U.S. Department of Energy, Office of Science, National Quantum Information Science Research Centers, Co-design Center for Quantum Advantage (C2QA) under contract number DE-SC0012704. We also acknowledge support from the DOE Early Career award (DE-SC0020120, for modeling of decoherence mechanisms and spin interactions), as well as AFOSR (FA9550-18-1-0334 and YIP FA9550-18-1-0081), the Eric and Wendy Schmidt Transformative Technology Fund, the Princeton Catalysis Initiative, and DARPA DRINQS (D18AC00015) for establishing the materials spectroscopy pipeline and developing integrated nanophotonic devices. We acknowledge the use of Princeton’s Imaging and Analysis Center, which is partially supported by the PCCM, an NSF MRSEC (DMR-1420541), as well as the Princeton Micro-Nano Fabrication Center.

\clearpage
\bibliographystyle{naturemag}
\bibliography{library.bib}
\clearpage

\clearpage

\section*{Methods}

\section{Experimental details}
\label{sec:experimental_setup}
In this work, we use a CaWO$_4$ sample (SurfaceNet GmbH) polished at (100) orientation (surface in a-c plane) with erbium introduced by ion implantation (II-VI Inc.) using an energy of 35 keV and a fluence of $5 \times 10^9$ ions per cm$^2$. Following the implantation, the sample was annealed in air at a temperature of 300~$^{\circ}$C for one hour. The nanophotonic devices stamped on the surface of the CaWO$_4$ sample were fabricated from silicon-on-insulator wafers following the procedure described in ref.~\cite{Chen2021a}. The cavity used in this work has a quality factor of $Q=1.6 \times 10^5$, allowing to Purcell enhance the Er$^{3+}$ emission rate, such that emission lifetime for the investigated ion is reduced from 6.3~ms to 18.4~$\mu$s for the optical $B$ transition, used for generating spin-photon entanglement. Due to its larger detuning from cavity resonance, the optical $A$ transition has a longer lifetime of 85.2~$\mu$s. The device is tuned on resonance with the Er$^{3+}$ optical transitions via nitrogen condensation. A photonic crystal grating coupler and an angle-polished single-mode fiber are used to couple light to the cavity. The device and sample are cooled to $T = 0.47$~K in a $^3$He cryostat (BlueFors LD250HE).

For optical addressing of Er$^{3+}$ transitions, we use a continuous-wave tuneable laser (Toptica CTL1500) offset locked to a second laser (Toptica CTL1500), which is frequency stabilized to ultra-low expansion reference cavity (Stable Laser Systems) by Pound–Drever–Hall lock technique. Optical pulses are generated using an intensity-modulating electro-optic modulator and two acousto-optic modulators to provide high extinction. Time-delayed fluorescence is detected using an SNSPD (Photon Spot). To avoid saturating the detector, the SNSPD bias current is turned off during optical excitation.

The spin transitions are driven with microwave pulses delivered by a wirebond antenna attached to the end of the optical fiber. Microwaves are generated using two IQ-modulated synthesizers (Agilent PSG E8267D) with signals being amplified to 3W before entering the cryostat.

Unless otherwise indicated, all experiments are performed at magnetic field strength of $|B| = 943.5$~G at orientation ($\theta,\phi$) = (85$^{\circ}$, -22$^{\circ}$), where $\theta$ is the angle from the \cawo{} $c$-axis and $\phi$ is the angle from the $a$-axis, which yields an optimal cyclicity for the optical $B$ transition of $C = 600(10)$.
At this magnetic field setting, we work with a ground state spin-splitting of $\omega_g = 10.7$~GHz and excited state spin-splitting of $\omega_e = 9.5$~GHz and drive the spin transitions with Rabi frequencies $\Omega_{R,g} = 13$~MHz and $\Omega_{R,e} = 7$~MHz respectively, for the ion in this work.  Finally, for the experiments probing optically induced spin dephasing, the optical pulse is detuned by 1~GHz from the ion.

\section{Optically induced spin dephasing}
\label{sec:op_ind_spin_dep}

As discussed in the main text, we observe that optical pulses induce dephasing of the spin in the form of random but static frequency fluctuations (Fig.~\ref{fig:Fig2}). Here, we supplement the discussion and report additional experiments to understand the origin of the dephasing mechanism.

First, we discuss the time-dependence of the dephasing effect. For optical pulses that are short with respect to the total evolution time, the effective evolution time $\tau_{e}$ under a random frequency shift, $\Delta\omega$ is given as $\tau_{e} = \min(\tau_{\textrm{opt}}, 2\tau-\tau_{\textrm{opt}})$, where $\tau_{\textrm{opt}}$ is the position of the optical pulse and $2\tau$ is the total evolution time of the Hahn sequence. The longest effective evolution time is achieved when the optical pulse is placed at the center of the Hahn sequence with $\tau_{e}=\tau$. For normally distributed $\Delta\omega$, the decay of the contrast over $\tau_{e}$ can be obtained as:
\begin{equation}
\label{eq:s_opt}
    S(\tau_{e}) = \int d(\Delta\omega) f(\Delta\omega,\sigma_{\omega}) \cos(\Delta\omega\tau_{e}) = 
    e^{-\frac{1}{2}(\sigma_{\omega}\tau_{e})^2},
\end{equation}
where we have integrated over a Gaussian frequency distribution, $f(\omega,\sigma_{\omega})$ with standard deviation $\sigma_\omega$. Using Eq.~\ref{eq:s_opt}, we infer $\sigma_{\omega}$, by fitting a Gaussian decay envelope with dephasing time $T_d = \sqrt{2}/\sigma_{\omega}$.

Next, we further investigate the dephasing dependence on optical pulse parameters. In Extended Data Fig.~\ref{fig:OptDep}a,b, we extract the dephasing magnitude $\sigma_{\omega}$ as a function of the optical power, $P$ and pulse-width, $W$. Before the observed saturation, fitting each to an exponential scaling of $\sigma_{\omega}\propto P^{\beta_P}$ and $\sigma_{\omega}\propto W^{\beta_W}$ yields exponents $\beta_P = 0.52(1)$ and $\beta_P = 0.43(6)$, consistent with a square root scaling with the number of photons, $N_P \propto P W $, so that
both experiments can be explained by a random walk model resulting in a frequency spread $\sigma_{P} = k_N \sqrt{N_P}$, where $k_N$ is an effective prefactor. 
To also account for the saturation observed, we fit to a simple model that limits the maximum frequency spread by the saturation value $\sigma_{\textrm{sat}}$ as $\sigma^2_{\omega} = \sigma_{\textrm{sat}}^2S^2/(S^2+1)$, where $S = \sigma_{P}/\sigma_{\textrm{sat}}$.
Such a saturation behavior could result from exhausting the configuration space of a finite sized bath. Comparing the extracted saturation, $\sigma_{\textrm{sat}}$ = $2\pi\times$370~kHz, to the long-term spin spectral diffusion $\sigma_{\omega}^*=\sqrt{2}/T_2^* = 2\pi\times$980~kHz, we find that the optically induced spin spectral diffusion may account for about 1/3 of the spin spectral diffusion over long time scales.

Finally, we investigate the origin of this dephasing mechanism, which could originate from magnetic or electric field noise. For \ercawo, electric fields are known to cause shifts of the \er{} g-tensor depending on the external magnetic field orientation \cite{mims1965electric}. In particular, the magnitude of the corresponding frequency shift, $\Delta \omega_E$, for a magnetic field in the aa-plane of the \cawo{} crystal is given as:
\begin{equation}
\label{eq:gshift}
    \Delta \omega_E  = \omega_g \left( \frac{\alpha\sin(2\phi-2\phi_0)}{2g_{\perp}^2}\right) E_z, 
\end{equation}
where $\omega_g$ is the frequency splitting of the ground spin state, $g_{\perp}$ is the g-factor in the aa-plane of \cawo{} and $E_z$ is the electric field along the \cawo{} $c$-axis. $\phi$ is the angle from the $a$-axis, $\phi_0 = 31^{\circ}$ and $\alpha = 11\cdot10^{-6}$~(V/cm)$^{-1}$, as discussed in Ref.~\cite{LeDantec2021} in the context of spin linewidths. In contrast to the varying sensitivity to electric field noise depending on the field orientation (Eq.~\ref{eq:gshift}), the magnetic moment in the aa-plane is uniform, suggesting that the sensitivity to magnetic field noise should not vary. In Fig.~\ref{fig:Fig2}e, we measure the optical dephasing effect at various magnetic field orientations on the aa-plane, which confirms the electric field origin of this effect. As expected by Eq.~\ref{eq:gshift}, the frequency spread varies as $\sigma_{\omega} = A |\sin(2\phi-2\phi_0^\prime)|$, where $\phi_0^\prime = 35(1)^{\circ}$ and $A = 2\pi\times103(4)$~kHz. The small difference between $\phi_0$ and measured $\phi_0^\prime$ could be due to errors in the crystal-cut or the magnetic field orientation. We perform the field sweep at $|B| = 600$~G, where the spin splitting is about $7$~GHz. Based on Eq.~\ref{eq:gshift}, this suggests an electric field fluctuation of $\sigma_{E_z}\sim 0.2$~kV/cm induced by the optical pulse in this experiment, equivalent in magnitude to a field from a single point charge at a distance of $\sim$100~nm. We note that the electric field fluctuations can be larger on a longer time-scale as inferred from the inhomogenous spin linewidth of a bulk ensemble in Ref.~\cite{LeDantec2021}. We note that the electric field origin of optically induced dephasing is also confirmed by a separate measurement of a $g=2$ ($S=1/2$) paramagnetic impurity coupled to the ion in this work. For the $g=2$ spin, we find that optical pulses have no effect on the spin coherence. Given the sensitivity of this impurity to magnetic noise, this agrees with our conclusion that the perturbation caused by the optical pulse is electric in origin rather than magnetic.

In the spin-photon entanglement experiments, we decouple this effect using the dynamical decoupling-based entanglement sequences. In Extended Data Fig.~\ref{fig:OptDep}c, we repeat the XY-16 experiment with an off-resonant optical pulse at a range of optical powers, which confirms that the sequence is robust to optically induced dephasing when the optical pulse is placed at a refocusing point of the sequence. While the dephasing mechanism can also be suppressed by an appropriate choice of the magnetic field orientation, to optimize cyclicity we are constrained to use a field orientation slightly off the aa-plane and at an angle that is sensitive to electric field fluctuations.
In future, the alignment between the photonic cavity and the crystal can also be adjusted to obtain a field orientation with good cyclicity that is also less susceptible to the optically induced dephasing effect.

\section{Spin coherence of the spin-photon state}
\label{sec:spin_coh}

Here, we discuss spin coherence during the spin-photon entanglement protocol with dynamical decoupling in the context of a generalized form of the protocol (Extended Data Fig.~\ref{fig:decoupling}).
Dynamical decoupling is a well studied technique to decouple fast noise \cite{biercuk2011dynamical}. Intuitively, our protocol simply performs dynamical decoupling in both the ground and excited spin manifolds to decouple noise in each, regardless of when the state decays from the excited state. However, the decay, at time $t$, can still occur in the middle of a decoupling unit, $(\tau-\pi-\tau)$, before the phase is entirely refocused in the excited manifold. Even in this case, we show that phase is still fully corrected for quasi-static noise. For fast noise, we find that there is a residual term, which can also be eliminated in the limit of smaller pulse spacings.

We consider decoupling under a dynamic noise bath, which can represent interactions with a bath of paramagnetic impurities, or charge noise as discussed in Methods~S\ref{sec:op_ind_spin_dep}, in a mean-field approach \cite{sousa2009electron}. In the rotating frame for the ground and excited spin manifolds, the 4-level Hamiltonian in the presence of dynamic noise can be given as:
\begin{equation}
    H_{\textrm{bath}} = \beta_g(t) S_{z,g} + \beta_e(t) S_{z,e},
\end{equation}
where $S_{z}^g = \frac{1}{2}\mathrm{diag}(0,0,1,-1)$ and $S_{z}^e= \frac{1}{2}\mathrm{diag}(1,-1,0,0)$ are the ground and excited spin operators respectively. For noise that is magnetic in origin, the ground and excited state noise terms are related by the magnetic moments of the spin states such that $\beta_e(t) = r\beta_g(t)$, where $r = \omega_e/\omega_g$. For noise originating from electric field fluctuations, the noise terms should still be correlated but the scaling, $r$, would depend on the distortion of the excited state g-tensor of \ercawo, which has not been measured to our knowledge.

We calculate the phase accumulated in our sequence under this noise model as represented in Extended Data Fig.~\ref{fig:decoupling}. A decoupling sequence with $N$ pulses can be expressed as $N$ repetitions of the decoupling unit as $(\tau-\pi-\tau)^N$. If spontaneous emission occurs on the boundary of a decoupling unit (\emph{i.e.} $t=2\tau k$, $k\in\mathcal{Z}$), the total phase can be simply expressed as a sum over decoupling units before and after the emission, equivalent to standard decoupling sequences. In general, as indicated by the highlighted areas in Extended Data Fig.~\ref{fig:decoupling}, when spontaneous emission does not occur at a boundary, a residual phase term will emerge:
\begin{equation}
\label{eq:res_phase}
    \Tilde{\phi}_{r} = \frac{1-r}{2}(2\tau k_e -t)\left(\Bar{\beta}_g(t,2\tau k_e) - \Bar{\beta}_g(T + t, T + 2\tau k_e)\right),
\end{equation}
where $k_e = \textrm{round}(t/2\tau)$ is the nearest decoupling unit to the emission time $t$ and $T$ is the time between the two optical pulses. $\Bar{\beta}_g(t_i,t_f)$ denotes time-average over the time-interval, $[t_i,t_f]$ defines as $\Bar{\beta}_g(t_i,t_f) =  \int_{t_i}^{t_f}\beta_g(t)dt/(t_f-t_i)$. First, we note that if the noise is quasi-static, constant over the duration of the entanglement attempt, the phase resulting from the decay time after the first excitation is cancelled by an equivalent term after the second excitation, after a time $T$. The cancellation of these terms after a longer time implies that noise at these two points is less likely to be correlated and may not be canceled efficiently. Second, the residual phase term is suppressed by correlations $r$ between the ground and excited state, disappearing for $r=1$. We expect that the optically induced dephasing mechanism discussed in Methods~S\ref{sec:op_ind_spin_dep} will induce an uncorrected frequency shift for this residual term, whose magnitude depends on the emission time $t$. In our experiment, we do not observe a strong dependence of the Bell state fidelity with respect to the emission time (Extended Data Fig.~\ref{fig:rate}), indicating that this residual term is small. Finally, the duration of the residual phase accumulation, $(2\tau k_e -t)$, can at most be $\tau$ so that $\Tilde{\phi}_{r}$ will disappear in the limit of choosing shorter $\tau$ for the decoupling sequence. 

\section{Optical decoherence during the entanglement sequence}
\label{sec:spin_photon_op}
In this section, we calculate the optical phase of the spin-photon state after the photon travels through the interferometer. Then, we briefly discuss the loss of coherence due to fast optical spectral diffusion and how we estimate the optical coherence in our experiment.

First, we discuss the regime where all optical frequency fluctuations are static over a single attempt and show that slow spectral diffusion does not lead to phase errors. As discussed in the main-text, a MW$_g$ $\pi/2$ pulse followed by an optical $\pi_B$ pulse prepares the superposition $(\ket{\downe}+\ket{\upg})/\sqrt{2}$. Spontaneous emission of this state in an even time-window yields $(e^{i(kx-\omega_B\tau_1)}\ket{\downg}\ket{E}+\ket{\upg}\ket{0})/\sqrt{2}$, where $\tau_1$ is the detection time after the optical $\pi_B$ pulse, $x$ is the optical path-length of the long arm, $\omega_B$ is the frequency of the probed transition and $k = \omega_B/c$ is the wavevector. After an odd number of $\pi$ pulses and a second optical excitation at time $T$, we obtain the state $(e^{i(kx-\omega_B\tau)}\ket{\upg}\ket{E}+e^{-i\omega_L T}\ket{\downe}\ket{0})/\sqrt{2}$, where the phase of the laser at frequency $\omega_L$ is imprinted on the excited term. Finally, the spontaneous emission of the second excitation and completion of the decoupling sequence prepares the spin-photon entangled state:
\begin{equation}
    \ket{\psi_{\textrm{Bell}}} = \frac{1}{\sqrt{2}}\big{(}e^{i(kx-\omega_B\tau_1)}\ket{\upg}\ket{E}+e^{-i\omega_L T}e^{-i\omega_B \tau_2}\ket{\downg}\ket{L}\big{)},
\end{equation}
where $\tau_2$ is the detection time after the second optical $\pi_B$ pulse. 
The detection of a single photon at time $\tau_{\textrm{det}}$, projects both $\tau_1$ and $\tau_2$ so that $\tau_{\textrm{det}} = \tau_1 = \tau_2 + T$. Each variable can also be broken down into a time of emission and propagation time, $\tau_1 = t_1 + T^{\prime}$ and $\tau_2 = t_2$, where $T^{\prime} = x/c$ is the true propagation delay of the interferometer and $t_{1(2)}$ is the time of emission after the first (second) optical excitation. After simplification, we find that the relative phase of the Bell state in the quasi-static regime is given as $\phi_{\textrm{Bell}} = \phi + \Delta\omega\Delta T$, where $\Delta\omega = \omega_B-\omega_L$ is the detuning of the emitter from the laser frequency, $\Delta T = T^{\prime} - T$ is the difference between the true propagation delay of the interferometer and the experimental delay and $\phi = \omega_L T^{\prime}$ is the interferometer phase at the laser frequency. For our parameter regime, the shot-to-shot spectral diffusion $\Delta\omega$ is on the order of 100~kHz and the propagation delay mismatch is on the order of 10 ns so that the phase error is very small ($\Delta\omega\Delta T \ll 1$).

However, optical frequency fluctuations within a single entanglement attempt can decohere the Bell state. 
To estimate the effect of fast frequency fluctuations on the optical coherence of the spin-photon entanglement experiment, we use the Hong-Ou-Mandel (HOM) visibility, which probes the indistinguishability of photons separated by the interferometer delay $T$ (Extended Data Fig.~\ref{fig:HOM}). For this estimate, we assume that frequency fluctuations are either much slower or much faster than the time-scale of our experiment. As argued above, any noise that is static over an entanglement attempt does not lead to optical dephasing. For the fast noise, we assume a Markovian noise model, where the frequency fluctuations are drawn from a Gaussian distribution with an infinitesimally short correlation time \cite{bylander2003interference}. Following the discussion in Ref.~\cite{ourari2023indistinguishable}, we fit the HOM interference to obtain a pure dephasing time $T_{\phi}$ under this model, leading to a reduced coherence $F(t) = \exp(-2t/T_{\phi})$ for a given emission time $t$, or equivalently a detection time-difference for the HOM experiment. The average optical coherence for photons collected in a given time-window, $[t_1,t_2]$ can then be calculated as:
\begin{equation}
    f_{op}(t_1,t_2) = \frac{\int_{t_1}^{t_2}e^{-t/T_1} F(t) dt}{\int_{t_1}^{t_2}e^{-t/T_1} dt }= 
    \frac{T_2}{2T_1}\frac{\exp\left(2t_1/T_2 \right) - \exp\left(2t_2/T_2 \right)}{\exp(t_1/T_1) - \exp(t_2/T_1) },
\end{equation}
where $1/T_2 = 1/2T_1 + 1/T_{\phi}$ and $T_1=\tau_r$. The HOM fit yields a dephasing time $T_{\phi} = 31.5$~$\mu$s, which leads to an $f_{op}(0,2.5\mu\textrm{s}) = 0.93$ for the time-window used in the Fig.~\ref{fig:Fig4}. In Extended Data Fig.~\ref{fig:spin_photon_XY20}, we extend the entanglement sequence to additionally use a later collection window, which is estimated to have a lower optical coherence of $f_{op}(7.5\mu\textrm{s},12.5\mu\textrm{s}) = 0.53$. We also note that, as discussed for spin coherence in the main-text, electric field fluctuations induced by the optical pulse also cause a frequency shift of the optical transitions. We do not analyze the effect of this on the optical coherence here.

\section{Phase monitoring}
\label{sec:phase_monitoring}

Here, we discuss the experimental setup and analysis to measure the phase of the interferometer. The phase of the MZI is tracked using a pair of avalanche photodiodes (APDs) placed in one of the output ports of the MZI (see Extended Data Fig.~\ref{fig:Phase}a), registering the interference of two reference laser pulses sent through the interferometer after each entanglement attempt. The APD signals are recorded using an oscilloscope (PicoScope 5000). The two phase tracking pulses are generated at the same time as the two $\pi_B$ pulses in the spin-photon entanglement experiment (Fig.~\ref{fig:Fig3}). To generate these pulses, a laser beam is divided into two optical paths (not shown), where a small fraction of laser light enters the cryostat for the optical excitation of the ions and a larger fraction is used for phase tracking to provide sufficient SNR. The phase tracking pulses are passed through an optical switch, turned on for the early and late entanglement attempt pulses only, before entering an input port of the MZI. Due to the scattering of the phase tracking laser pulses in the 15.6 km fiber spool of MZI, an additional time-dependent signal is observed in the time trace of the background counts as shown in Extended Data Fig.~\ref{fig:Bck}a. To avoid the overlap of the entangled photons with this scattering signal, phase monitoring pulses are delayed by 7.3~$\mu$s with an additional 1.5~km fiber spool before entering the MZI. For the experiment in Extended Data Fig.~\ref{fig:spin_photon_XY20}, a 3~km fiber spool is used instead to delay the scattering further and avoid the overlap with the later photon collection window.

Next, we discuss the required analysis to obtain the phase. The two phase tracking pulses sent through the interferometer accumulate a relative phase, $\phi$. When recombined, this relative phase results in amplitudes of $1\pm\cos\phi$ in the two output ports of the interferometer. By measuring one of these output amplitudes using an APD, the phase $\phi$ can be measured up to an integer multiple of $\pi$.
To resolve the full phase up to $2\pi$, we use a technique developed in Ref.~\cite{zheng2024reducing} based on simultaneous measurements of interference that are out-of-phase. Here, the birefringence property of the 15.6 km long optical fiber is utilized, which is induced by bending strain in the fiber spool. Laser pulses polarized along the fast axis or slow axis can gain different phases, i.e. $\mathbf{E}_{fast}\rightarrow e^{i\phi}\mathbf{E}_{fast}$ and $\mathbf{E}_{slow}\rightarrow e^{i(\phi+\delta\phi)}\mathbf{E}_{slow}$, corresponding to two different amplitudes after the interferometer, $1+\cos\phi$ and $1+\cos(\phi+\delta\phi)$. Because the phase difference $\delta\phi$ is random and usually non-zero, $\cos\phi$ and $\cos(\phi+\delta\phi)$ do not reach maximum or minimum at the same time, leading to reduced interferometer visibility if the input polarization is not aligned with one of the principal axes. 
To align the birefringent axes with the polarization of single photons, equivalently the cavity polarization, we optimize the interference contrast of the cavity reflected light on SNSPDs after the MZI.
After traveling through the MZI, the phase tracking pulses are split using a PBS to separately read the phase of the fast and slow axes polarizations.
Thus, each phase monitoring yields two measurements, $\cos\phi$ for the polarization of the single photons, and $\cos(\phi+\delta\phi)$ for the polarization perpendicular to the single photons, revealing an ellipse when plotted against each other over many trials (Extended Data Fig.~\ref{fig:Phase}b). The eccentricity of the ellipse can be fitted to extract $\delta\phi$ and determine the phase, $\phi$, of each point on the ellipse up to an integer multiple of $2\pi$. The phase difference $\delta\phi$ changes by less than a degree over twenty minutes, while the phase $\phi$ of the interferometer only remains stable for several milliseconds, as shown in Extended Data Fig.~\ref{fig:Phase}c,d.

The knowledge of $\cos\phi$ and $\cos(\phi+\delta\phi)$ can resolve the phase up to $2\pi$, but cannot distinguish $\pm\phi$ without knowing the sign of $\delta\phi$, i.e., which axis is fast or slow. In this work, we do not distinguish the sign, but instead keep track of $\delta\phi$ such that it does not cross 0 or $2\pi$.

\section{Conditional operation}
\label{sec:cond_op}
To increase the spin-photon entanglement generation rate, we perform a spin readout conditional on the detection of a single photon in the heralding window. For that purpose, we utilize a complex programmable logic device (CPLD), which switches the experiment state between two modes: (E) initialization followed by the entanglement attempt, and (R) spin readout (see Fig.~\ref{fig:Fig4}a in the main text). The default CPLD state is E, so that entanglement attempts are performed most of the time. In the case of relatively rare events of single photon detection (probability of $6.7 \times 10^{-4}$ per attempt), the CPLD changes the experiment state to R. After sending 432 optical readout pulses (lasting around 49~ms), it switches back to E mode. The CPLD switches the experiment between the two modes by controlling the flow of the gating and triggering signals to the switches and waveform generators responsible for optical and MW pulse generation. The basic block sequence in our experiment is defined by 6 periods each lasting 75.5~$\mu$s. In E mode, it consists of 2 initializing pulse pairs [Methods~S\ref{sec:init_and_readout}], an empty period followed by an entanglement attempt consisting of 2 optical pulses, and ends with another empty period. Entanglement pulses are sandwiched between empty periods to prevent overlap of entanglement photons with photons emitted after the last initialization and the first readout pulses (due to the delay picked up in the MZI). In R mode, the block sequence is repeated 108 times and consists of four readout pulses and two empty periods. The presence of empty periods in R mode is inherited from the E sequence, but it does not serve any purpose. Conditional spin readout allowed us to increase the spin-photon entanglement generation rate by two orders of magnitude by skipping the slow readout when no photon is detected within the heralding window.

\section{Initialization and readout}
\label{sec:init_and_readout}
For spin initialization, we used a sequence consisting of optical $\pi_A$ pulses, each followed by a chirped microwave pulse resonant with an excited state spin transition. The chirped pulses were used to achieve a high inversion of the excited state spin population, in the presence of interaction with two paramagnetic impurities with interaction strength near 3.3~MHz, close to the Rabi frequency of $\Omega_{R,e} = 7$~MHz for the excited state.
This allowed us to initialize using only two initialization cycles in the spin-photon entanglement experiment with an estimated fidelity of 98.5\% based on fluoresence measured after initialization.

For spin readout, we used a sequence of 240 and 432 optical $\pi_B$ pulses each followed by a fluorescence collection window of 70~$\mu$s, for spin dynamics experiments without the MZI and spin-photon entanglement experiments, respectively. To differentiate the two spin states, a threshold of one photon is used. The number of readout pulses for each type of experiment is optimized to maximize the readout fidelity. For spin dynamics experiments, after initializing the spin into the $\ket{\uparrow}$ ($\ket{\downarrow}$) state, an average of 3.1 (0.1) photons is observed, allowing for $\ket{\uparrow}$ ($\ket{\downarrow}$) readout with fidelity $F_{\uparrow}=0.93$  ($F_{\downarrow}=0.85$). For the spin-photon entanglement experiment, the spin-readout fluorescence signal is passed through the MZI with a 15.6 km fiber delay, introducing 72\% of additional photon loss. In this case, an average of 1.4 (0.2) photons is observed after initializing spin into the $\ket{\uparrow}$ ($\ket{\downarrow}$) state, allowing for $\ket{\uparrow}$ ($\ket{\downarrow}$) readout with fidelity $F_{\uparrow}=0.81$  ($F_{\downarrow}=0.69$). Corresponding histograms of photon counts are shown in Extended Data Fig.~\ref{fig:SingleShot}. We note that both the spin dynamics and spin-photon entanglement data shown in the main manuscript are normalized to readout fidelity. The readout fidelity here is lower than in previous work \cite{ourari2023indistinguishable}, primarily because of the lower Purcell factor, resulting in lower cyclicity and higher dark counts accumulated over a longer photon collection window. For entanglement experiments an additional factor is photon loss in the MZI, which can be mitigated in the future by bypassing the MZI with an optical switch during spin readout.

\section{Spin-photon entanglement fidelity estimation}
\label{sec:fidelity_estimation}
Here, we describe the procedure to estimate the spin-photon entanglement fidelity in each basis.
For ZZ basis measurements, which correlate the spin state in the $Z_s$ basis and the photon state in the $Z_p$ basis, the correlation between spin and photon measurements can be directly calculated, as shown in Fig.~\ref{fig:Fig4}e. For XX (YY) basis measurements, which correlate the spin state in the $X_s$ ($Y_s$) basis and the photon state in the $\Phi_p$ basis, we use Bayesian inference to estimate the finite fidelity spin-photon state most likely to yield the measurement results.
Suppose the state is in the form of $\rho(\alpha)=\alpha\ket{\Psi}\bra{\Psi}+(1-\alpha)I$, where $\ket{\Psi}$ is the desired spin-photon Bell state and $I$ is the fully mixed state. Projecting $\rho(\alpha)$ onto the measured spin and photon state gives the conditional probability to obtain the measurement given $\alpha$: $P(\ket{\uparrow,\pm\phi}|\alpha)=(1\pm\alpha\cos\phi)/2$ for XX basis and $P(\ket{\uparrow,\pm\phi}|\alpha)=(1\pm\alpha\sin\phi)/2$ for YY basis.
The likelihood of parameter $\alpha$ given the dataset is $P(\text{result}|\alpha)=\Pi_{i}P(\ket{S_{i},P_{i}}|\alpha)$, taking a product over all heralded entanglement attempts indexed by $i$.
Finally, the likelihood function $f(\alpha)=P(\text{result}|\alpha)$ is fitted using a Gaussian to extract the value of $\alpha$ that is most likely to reproduce the result and its uncertainty. The contrast of the oscillation for the XX or YY basis is given by the fitted parameter $E_{X/Y}=\alpha^{*}$. While data is not binned for the estimation, we bin it into 11 phase points to plot it in Fig.~\ref{fig:Fig4}c,d.

We correct the spin-photon correlations for the readout fidelity after fitting. 
The fidelities of reading two spin states are $F_{\uparrow}=0.81$ and $F_{\downarrow}=0.69$ respectively  [Methods~S\ref{sec:init_and_readout}]. Therefore, the probability of measuring state $\ket{\uparrow}$ is $P(m=\ket{\uparrow})=F_{\uparrow}P(\ket{\uparrow})+(1-F_{\downarrow})P(\ket{\downarrow})$. Alternatively, given the measured probability, the readout corrected population is $P(\ket{\uparrow})=(P(m=\ket{\uparrow})-1+F_{\downarrow})/(F_{\uparrow}+F_{\downarrow}-1)$. Thus, the measured contrasts in the spin-photon entanglement experiment are scaled by $(F_{\uparrow}+F_{\downarrow}-1)^{-1}$ to obtain the readout corrected fidelity.

\section{Extended spin-photon sequence}
\label{sec:extended_spin_photon}

As mentioned in the main-text, we limit our photon collection window to the first 2.5~$\mu$s for the XY-16 based entanglement sequence to minimize sensitivity to fast dephasing of the optical transition.
Here, we discuss an additional spin-photon entanglement experiment based on the XY-20 dynamical decoupling sequence, which allows us to use a second collection window by prolonging the sequence duration to include this emission window (Extended Data Fig.~\ref{fig:spin_photon_XY20}a,b). Then, we briefly discuss requirements to include all emitted photons in the sequence.

In Extended Data Fig.~\ref{fig:spin_photon_XY20}c,d, we observe parity oscillations of the Bell state for both the first and second heralding windows. Photons collected from the two windows yield visibilities of $\{E_{X_1},E_{Y_1},E_{Z_1}\} = \{0.48(3), 0.47(3), 0.80(1)\}$ and $\{E_{X_2},E_{Y_2},E_{Z_2}\} = \{0.14(3), 0.09(3), 0.72(1)\}$, leading to fidelities $F_1 = 0.69(1)$ and $F_2 = 0.49(1)$ respectively. The slightly lower visibility of parity oscillations in the first window for this experiment in comparison to the XY-16 based experiment is due to lower spin coherence ($f_s = 0.67$) after the longer spin evolution time under the XY-20 sequence. 
The lower visibility of the second window is primarily due to the lower optical coherence for cases where spontaneous emission occurs in the second heralding window of [7.5~$\mu$s -- 12.5~$\mu$s] ($f_{op} = 0.53$), which can be roughly estimated based on the dephasing time extracted from the HOM experiment [Methods~S\ref{sec:spin_photon_op}]. 
Another factor for the lower fidelity is the higher impact of background counts due to the lower emission rate in the second window ($f_{bg(X/Y)}=0.91$, $f_{bg(Z)}=0.92$).
A final factor may be larger pulses errors experienced in the excited state arising from the lower Rabi frequency ($\Omega_{R,e}=7$~MHz), comparable to interaction strength of 3.3~MHz with two paramagnetic impurities.
While the second window has lower fidelity, integrating over all photons in both windows, we obtain a rate of $R = 2.8$~Hz and visibilities of $\{E_X,E_Y,E_Z\} = \{0.31(2), 0.30(2), 0.76(1)\}$, leading to an entanglement fidelity of $F = 0.59(1)$, still above the classical threshold.

To make use of all emitted photons, we need larger optical and spin coherence and additionally track the interferometer phase at the frequency $\omega_A$, corresponding to photons emitted in the odd windows.

\section{Error model for spin-photon entanglement}
\label{sec:error_model}
The error sources of the spin-photon entanglement generation in this work (Fig.~\ref{fig:Fig4}) mainly consist of finite optical and spin coherence, pulse errors, background photon counts, initialization fidelity, and finite optical lifetime. The error values, $\epsilon$, stated in the main-text are expressed as fidelity contributions, $f = 1 - \epsilon$, in the following discussion.

The optical coherence can be obtained from the Hong–Ou–Mandel (HOM) visibility through the same interferometer used in the experiment, as shown in Extended Data Fig.~\ref{fig:HOM}. The end of the heralding window of 2.5 $\mu$s after the optical excitation used in the entanglement experiment gives a HOM visibility of $f_{op}=0.93$. The spin coherence during the spin-photon entanglement experiment is assumed to be similar to the spin coherence under XY-16 dynamical decoupling sequence because the ion spends at most 2.5 $\mu$s in the excited state compared to the total free evolution time of 80.5 $\mu$s. The heralding measurement also ensures that both spin states spend the same amount of time in the excited state such that the extra phases acquired due to excitation are canceled by the dynamical decoupling as discussed in Methods~S\ref{sec:spin_coh}. Furthermore, the standard XY-16 sequence contains errors due to imperfect pulses. Based on Extended Data Fig.~\ref{fig:XY16}, the spin coherence is $f_{s}=0.75$ after 80.5 $\mu$s of total evolution time used in the entanglement experiment. Both the optical and spin coherence induced errors contribute only to XX and YY basis in the spin-photon entanglement. For the ZZ basis, optical coherence is assumed to be irrelevant. To estimate the effect of pulse errors, we apply the XY-16 decoupling sequence, without $\pi/2$ pulses, directly on the $\ket{\upg}$ and $\ket{\downg}$ states, which yields a reduced contrast of $f_{p}=0.86$.

Dark counts of the SNSPDs and the laser scattering contribute to the background photon counts we observe in the experiment. The background is measured by repeating the spin-photon entanglement experiment sequence with the laser detuned from the ion by 100 MHz, as shown in Extended Data Fig.~\ref{fig:Bck}. In the ZZ basis, the background photon counts are similar to the dark count rate of 6 Hz, and the two heralding windows yield a false heralding probability of $4\%$, or $f_{bg(Z)}=0.96$. In the XX/YY basis, the background counts are higher than the dark counts at around 15 Hz due to scattered photons from the phase tracking pulses. The single XX/YY heralding window yields a false heralding probability of $5\%$, or $f_{bg(X/Y)}=0.95$.

Initialization errors also reduce the contrast of the Bell state parity oscillations. A false initialization probability of 1.5\% reduces the contrast of parity oscillations by 3\%, so that the contribution is $f_{i}=0.97$. However, initialization errors do not affect the ZZ basis measurement since the heralding of the $\ket{E}$ or $\ket{L}$ photon state correctly projects the spin-state regardless of initialization errors.

The finite optical lifetime can lead to a residue excited state population excited by the first optical pulse, which does not decay before the second optical excitation. During the spin-photon entanglement sequence, the probabilities of photon emission during the second heralding window excited by the first or second optical pulses are $P_{1}=e^{-T(1/2\tau_{A}+1/2\tau_{B})}(1-e^{-\delta/\tau_{A}})=0.2\%$ and $P_{2}=1-e^{-\delta/\tau_{B}}=9.8\%$ respectively. Here $\delta=1.9$ $\mu$s is the heralding window width, $T=75.5$ $\mu$s is the separation between two optical pulses, and $\tau_{A}=85.2$ $\mu$s and $\tau_{B}=18.4$ $\mu$s are the optical lifetime of two optical transitions of the ion. This corresponds to the contrast of $f_{l}=P_2/(P_1+P_2)=0.98$ for both ZZ and XX/YY basis measurements.

By accounting for the above source of errors, the expected contrasts are $E_{X/Y}=f_{op}f_{s}f_{l}f_{bg(X/Y)}f_{i}=0.63$ and $E_{Z}=f_{p}f_{l}f_{bg(Z)}=0.81$, compared to the experiment result $E_{X}=0.60$, $E_{Y}=0.55$, and $E_{Z}=0.77$. The expected final fidelity is 0.78, compared to the experiment result of 0.73.

\section{Entanglement rate}
\label{sec:entanglement_rate}
In this section, we discuss the current entanglement rate and consider its extensions. We start by discussing factors yielding the current rate and estimate the spin-photon entanglement rate that can be achieved with improvements to these factors.

For the spin-photon entanglement generation experiment, we use two initialization pulse pairs followed by two optical excitation pulses (Fig.~\ref{fig:Fig4}a), repeated at a rate of 2.2 kHz. 
The entanglement generation is successful with probability $P_{\textrm{ent}}=6.7 \times 10^{-4}$, leading to the entanglement rate of 1.48~Hz in our experiment, which is not limited by the readout performed conditionally [Methods~S\ref{sec:cond_op}].
The probability $P_{\textrm{ent}}$ is limited by the total photon collection efficiency and the finite width of the photon collection window in our protocol. The experimentally recorded probability of detecting a single photon on an SNSPD after optical excitation is around $0.02$, which is limited by several factors, including the photon extraction efficiency from the cavity $\eta_{cav} = 0.24$, the grating coupler efficiency $\eta_{gc} = 0.33$, the transmission through the passive optical components $\eta_{net} = 0.61$ and the SNSPD detection efficiency $\eta_{det} = 0.85$.
The combined predicted detection probability is $P=\eta_{cav} \times \eta_{gc} \times \eta_{net} \times \eta_{det} = 0.04$. For spin-photon entanglement experiments, efficiency is further decreased by the losses of the MZI $\eta_{MZI} = 0.28$ (dominated by 15.6 km fiber delay) and the narrow photon collection window of 1.9 $\mu$s, corresponding to $\eta_{pc} = 0.091$. This gives an overall predicted entanglement generation probability of $P_{\textrm{ent}} = P \times \eta_{MZI} \times \eta_{pc} = 1.0 \times 10^{-3}$, close to the measured value of $6.7 \times 10^{-4}$. 

The rate can be further improved with straightforward improvements to these parameters. 
First, the entanglement attempt rate can be increased to 3.3 kHz (x1.5), for a total attempt duration of $(1 + 2 + 1)\times75.5$~$\mu$s, by using a single initialization pulse, two pulses for the entanglement attempt and an additional waiting period to collect all photons after the MZI. 
Using a cavity that is critically coupled would lead to an $\eta_{cav}$ of $0.5$ (x2), gaining in coupling efficiency without significantly reducing the $Q$ of the cavity. Switching to an adiabatically tapered fiber for coupling to the photonic device, instead of a grating coupler, could lead to a coupling efficiency of 0.97 (x3) \cite{tiecke2015efficient}.
Finally, using all emitted photons would lead to $\eta_{pc}$ of 1 (x11). As discussed in Methods~S\ref{sec:extended_spin_photon}, the use of all emitted photons is currently limited by fast optical dephasing and spin coherence errors. Improving optical and spin coherence would facilitate the collection of all emitted photons. Leaving other factors fixed, these improvements would lead to an entanglement rate of 150 Hz after 15.6 km.

\setcounter{figure}{0}
\renewcommand{\thetable}{\textbf{\arabic{table}}}%
\renewcommand{\figurename}{\textbf{Extended Data Fig.}}
\renewcommand{\tablename}{\textbf{Extended Data Table}}
\renewcommand{\thefigure}{\textbf{\arabic{figure}}}%
\clearpage
\section*{Extended Data}

\begin{figure*}[ht]
	\centering
    \includegraphics[]{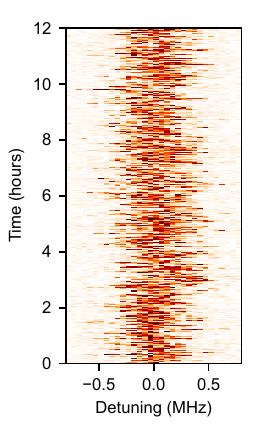}
    \caption{\textbf{Long term spectral stability.} \textbf{a} Repeated PLE scans of the investigated ion, with a 12 h averaged full-width, half-maximum linewidth of 470~kHz and an average single-scan linewidth of 240~kHz.
    }
    \label{fig:SD}
\end{figure*}

\begin{figure*}[ht]
	\centering
    \includegraphics[]{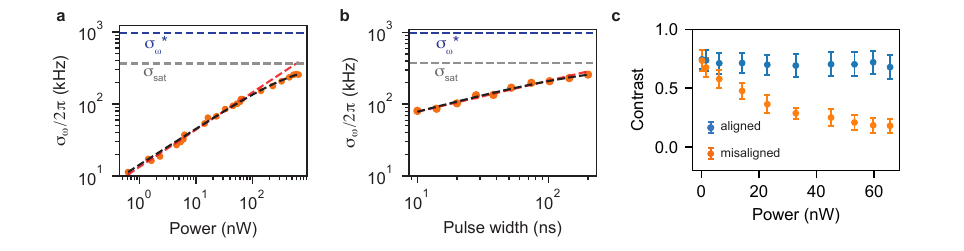}
    \caption{\textbf{Optically induced spin dephasing.} \textbf{a} Magnitude of spin spectral diffusion extracted as a function of optical power ($W = 200$~ns) fitted to a random walk model (black dashed line) with saturation $\sigma_{\textrm{sat}}$ = $2\pi\times$364(10)~kHz, which we compare to $\sigma_{\omega}^*=\sqrt{2}/T_2^* = 2\pi\times$980~kHz. Fitting $\sigma_{\omega} = k_PP^{\beta_P}$ (red dashed line) to the points before saturation ($< 100$~nW) yields an exponent of $\beta_P = 0.52$(1).
    \textbf{b} Magnitude of spin spectral diffusion extracted as a function of pulse-width ($P=655$ nW), fitted to the same model with saturation $\sigma_{\textrm{sat}}$ = $2\pi\times$370(23)~kHz. Fitting $\sigma_{\omega} = k_WW^{\beta_W}$ to points before saturation ($< 50$~ns) yields an exponent of $\beta_W = 0.43$(6). 
    \textbf{b} XY-16 spin coherence for a fixed total evolution time ($2N\tau=80.5$ $\mu$s) corresponding to the entanglement sequence when an off-resonant optical pulse ($w_p=200$ ns) is aligned (blue) or misaligned (orange). 
    The aligned case corresponds to the same setting for the spin-photon experiment, where the optical pulse is placed at a refocusing point of the sequence. In the misaligned case, the optical pulse is instead overlapped with a $\pi$ pulse of the decoupling sequence. No power dependence is observed when the optical pulse is properly aligned. When the pulse is misaligned, increasing optical power decoheres the spin as expected. Error bars correspond to one s.d.
    }
    \label{fig:OptDep}
\end{figure*}

\begin{figure*}[ht]
	\centering
    \includegraphics[]{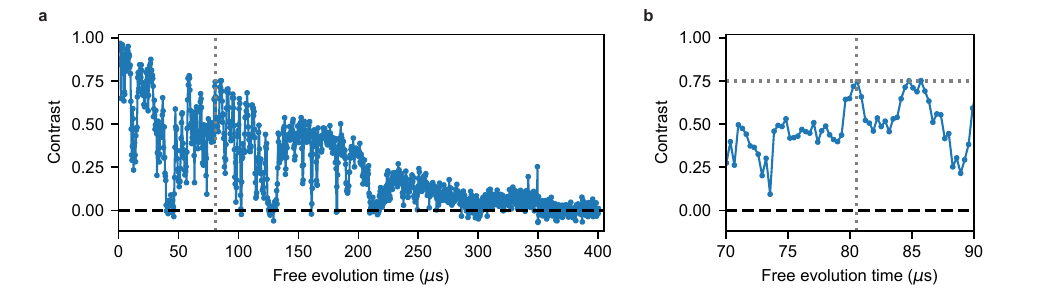}
    \caption{\textbf{Spin coherence under XY-16 sequence.} \textbf{a} The coherence of the ground spin state as a function of the total evolution time $32\tau$ during XY-16 dynamical decoupling sequence under a 1.6~nW off-resonant optical pulse equivalent to Fig.~\ref{fig:Fig3}d. The contrast is corrected for single-shot readout fidelity. \textbf{b} Zoom-in around evolution time corresponding to $\tau$ used in spin-photon entanglement experiment. }
    \label{fig:XY16}
\end{figure*}

\begin{figure*}[ht]
	\centering
    \includegraphics[]{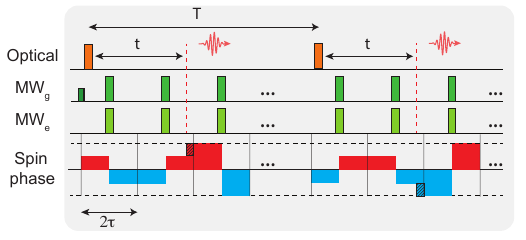}
    \caption{\textbf{Spin-photon decoupling sequence.} Spin phase accumulation during the spin-photon entanglement with dynamical decoupling for with inter-pulse spacing $2\tau$ and seperation between optical pulses of $T = 2\tau k_p$, where $k_p$ is an odd integer. The red dashed lines indicates the emission time of the photon after time $t$ with respect to each optical pulse. For slow noise, the phase is fully canceled, indicated by the total area summing to 0. The highlighted squares indicate the small amount of phase that is cancelled after a longer time $T$.}
    \label{fig:decoupling}
\end{figure*}

\begin{figure*}[ht]
	\centering
    \includegraphics[]{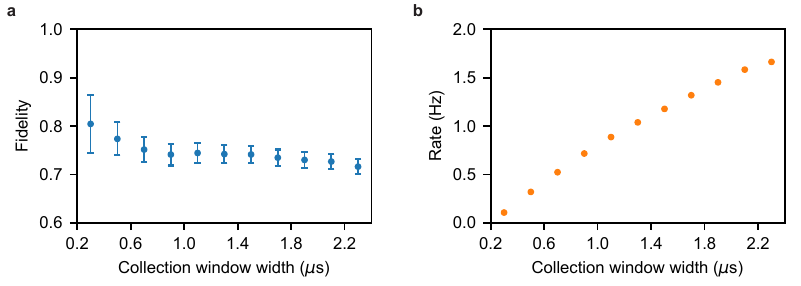}
    \caption{\textbf{Entanglement fidelity vs rate.} \textbf{a} The spin-photon entangled state fidelity and \textbf{b} rate as a function of the photon collection window width. The collection window starts 0.4~$\mu$s after the falling edge of the optical $\pi$ pulse. Error bars correspond to one s.d.}
    \label{fig:rate}
\end{figure*}

\begin{figure*}[ht]
	\centering
    \includegraphics[]{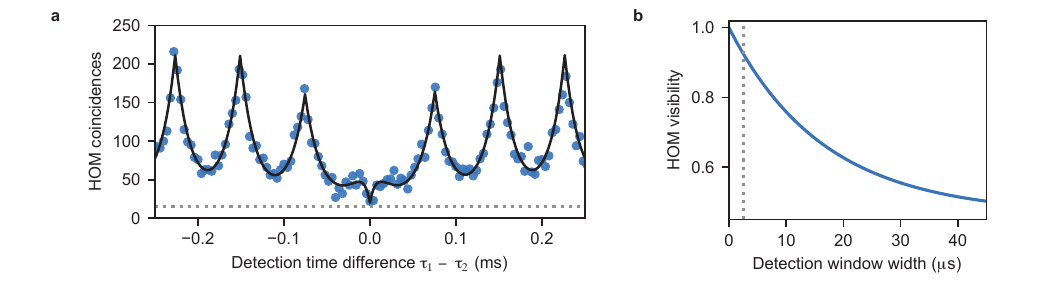}
    \caption{\textbf{Generation of indistinguishable photons.} \textbf{a} Two-photon interference histogram for photons emitted by the same ion spaced by 75.5~$\mu$s. The grey dashed line corresponds to an independently recorded background level. The solid black line is a fit. \textbf{b} HOM visibility extracted from the fit corrected for background counts and imperfections of the MZI. For the detection window width of 2.5 $\mu$s (corresponding to the photon detection window in the entanglement experiment), we observe a HOM visibility of 0.93.}
    \label{fig:HOM}
\end{figure*}

\begin{figure*}[ht]
	\centering
    \includegraphics[]{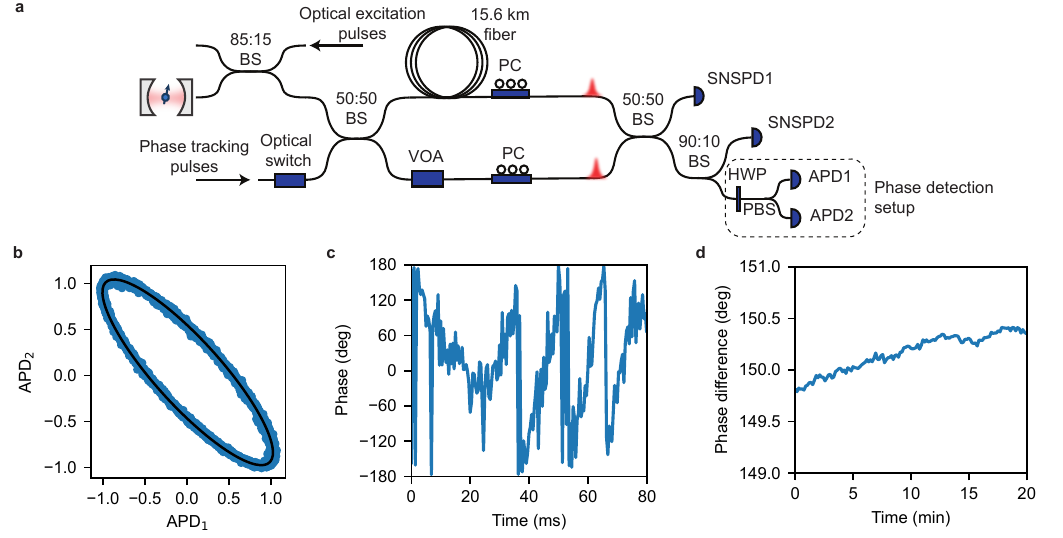}
    \caption{\textbf{Phase tracking.} 
    \textbf{a} Optical setup used to measure time-bin photons after 15.6~km of fiber in an unbalanced MZI and track the interferometer phase. Optical pulses for excitation and phase tracking are sent to nanophotonic cavity and interferometer respectively. Single photons from the \er{} ion are detected at two SNSPDs after the MZI. A variable optical attenuator (VOA) attenuates the short arm to match the attenuation rate of the 15.6~km fiber spool and polarization controllers (PC) are used to match the polarization of single-photons from each arm. Bright phase tracking pulses are measured in two orthogonal polarizations on avalanche photodiodes (APD), using a half-wave-plate (HWP) followed by a polarization beam splitter (PBS).
    \textbf{b} Ellipse constructed from the phase detection setup. 5000 data points measured in 2 seconds are shown to form the shape of an ellipse, with the fitted parametric equation $\text{APD}_1=\cos\phi$ and $\text{APD}_2=\cos(\phi+\delta\phi)$ with $0<\phi\le 2\pi$ and the fitting parameter $\delta\phi$. The location of each data point on the ellipse corresponds to the interferometer phase at the time of each measurement.
    \textbf{c,d} Extracted interferometer phase, $\phi$, (c) varies in the timescale of milliseconds and phase difference, $\delta\phi$, between the two polarization components (d) varies by less than a degree over 20 minutes.}
    \label{fig:Phase}
\end{figure*}

\begin{figure*}[ht]
	\centering
    \includegraphics[]{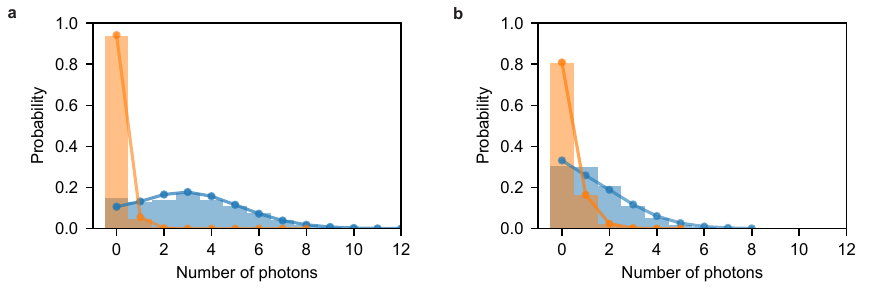}
    \caption{\textbf{Single shot spin readout.} Histograms of photon counts after initializing ion in spin up (orange) and spin down (blue) state. Solid lines are fit to a Poisson distribution model taking into account the finite cyclicity of the optical transition \cite{ourari2023indistinguishable}. \textbf{a} Histogram obtained for photons bypassing MZI (Hahn, XY16 experiments) using 240 readout pulses with average readout fidelity $F=89.4\%$. \textbf{b} Histogram obtained for photons passing through MZI (spin-photon entanglement experiment) using 432 readout pulses with average readout fidelity $F=75.1\%$.  In both experiments, we use a readout threshold of one photon.}
    \label{fig:SingleShot}
\end{figure*}

\begin{figure*}[ht]
	\centering
    \includegraphics[]{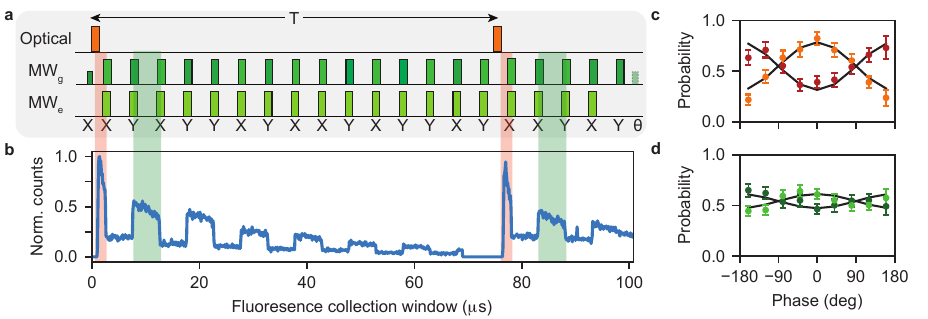}
    \caption{\textbf{Spin-photon entanglement with XY-20.} \textbf{a} An extended spin-photon entanglement sequence based on XY-20, which allows for two heralding windows, highlighted in orange and green. \textbf{b} Fluoresence corresponding to the pulse sequence. \textbf{c,d} Bell state parity oscillations for spin measurement in the $X_s$ basis for the first heralding window (c) and the second heralding window (d) with visibilities of $E_{X,1} = 0.48(3)$ and $E_{X,2} = 0.14(3)$ respectively. The lighter (darker) color corresponds to measuring the photonic qubit in $\ket{+\phi}$ ($\ket{-\phi}$).
    }
    \label{fig:spin_photon_XY20}
\end{figure*}

\begin{figure*}[ht]
	\centering
    \includegraphics[]{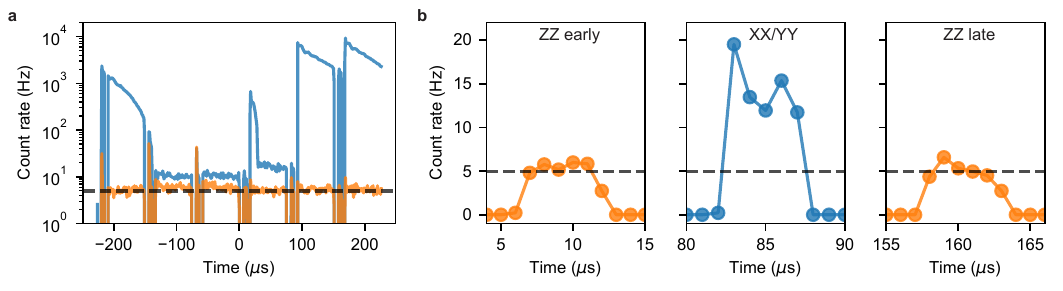}
    \caption{\textbf{Dark counts and laser scattering.} \textbf{a} Time-trace of background counts for spin-photon entanglement experiment with laser detuned by 100~MHz from the ion transition. The blue trace corresponds to the XX/YY basis experiment, where phase-tracking laser pulses are sent through the MZI, while the orange trace corresponds to the ZZ basis experiment without phase-tracking pulses. The phase-tracking laser pulses are delayed by 7.3~$\mu$s with respect to ion excitation pulses and lead to double scattering events in 15.6 km long fiber, visible as a significant increase of count rate reaching $10^3-10^4$ Hz. The dashed line corresponds to the total SNSPDs dark count level. \textbf{b} Zoom-in around the photon collection window showing average background counts on the level of 6 and 15 Hz for the XX/YY and ZZ basis experiments, respectively.}
    \label{fig:Bck}
\end{figure*}

\end{document}